\documentclass[prd,showpacs,nofootinbib,showkeys]{revtex4}

\usepackage{graphicx}
\usepackage{amsmath}
\usepackage{amssymb}
\usepackage{amsxtra}

\begin{document}

\title{Neutrino masses and mixing within a $SU(3)$ family symmetry model with one or two light sterile neutrinos}

\author{Albino Hern\'andez-Galeana}

\email{albino@esfm.ipn.mx}

\affiliation{ Departamento de F\'{\i}sica,   Escuela Superior de
F\'{\i}sica y Matem\'aticas, I.P.N., \\
U. P. "Adolfo L\'opez Mateos". C. P. 07738, M\'exico, D.F.,
M\'exico. }



\begin{abstract}
We report a global fit of parameters for fermion masses and mixing,
including light sterile neutrinos, within a local vector $SU(3)$ family symmetry model.
In this scenario, ordinary heavy fermions, top and bottom quarks and tau
lepton, become massive at tree level from {\bf Dirac See-saw} mechanisms
implemented by the introduction of a new set of $SU(2)_L$ weak singlet vector-like
fermions, $U,D,E,N$, with $N$ a sterile neutrino. The $N_{L,R}$ sterile neutrinos allow
the implementation of a $8\times 8$ general tree level See-saw Majorana neutrino mass
matrix with four massless eigenvalues. Hence, light fermions, including
light neutrinos obtain masses from one loop radiative corrections mediated
by the massive $SU(3)$ gauge bosons.  This BSM model is able to accommodate the
known spectrum of quark masses and mixing in a $4\times 4$
non-unitary $V_{CKM}$ as well as the charged lepton masses. The explored parameter
space region provide the vector-like fermion masses:
$M_D \approx 914.365\,$GeV, $M_U \approx 1.5 \,$TeV, $M_E \approx 5.97 \,$TeV, $SU(3)$
family gauge boson masses of $\mathcal{O} (1-10)\,$TeV, the neutrino masses
$(m_1, m_2, m_3, m_4, m_5, m_6, m_7, m_8)=
(0,0.0085, 0.049, 0.22, 3.21 , 1749.96,1\times 10^8, 1\times 10^9\,)\,$eV, with the
squared neutrino mass differences: $m_2^2-m_1^2 \approx 7.23 \times 10^{-5}\;\text{eV}^2$,
$m_3^2-m_1^2 \approx 2.4 \times 10^{-3}\;\text{eV}^2$, $m_4^2-m_1^2 \approx 0.049\;\text{eV}^2$,
$m_5^2-m_1^2 \approx 10.3\;\text{eV}^2$. We also show the corresponding $U_{PMNS}$ lepton mixing matrix.
However, the neutrino mixing angles are extremely sensitive to parameter space region, and
an improved and detailed analysis is in progress.
\end{abstract}

\maketitle

\tableofcontents


\section{ Introduction }
The standard picture of three flavor neutrinos has been successful to account
for most of the neutrino oscillation data. However, several experiments have reported
new experimental results, on neutrino mixing\cite{Altarelli.Smirnov}, on large $\theta_{13}$ mixing from Daya Bay\cite{DayaBay}, T2K\cite{T2K}, MINOS\cite{MINOS}, DOUBLE CHOOZ\cite{DOUBLE}, and RENO\cite{RENO}, implying
a deviation from TBM\cite{TBM} scenario. In addition, the recent experimental
results from the LSND and MiniBooNe short-baseline neutrino oscillation experiments,
provide indications in favor of the existence of light sterile neutrinos in the eV-scale, in order
to explain the tension in the interpretation of these data\cite{MiniBooNE,LSND-MiniBooNe}.

\vspace{3mm}
The strong hierarchy of quark and charged lepton masses and quark mixing
have suggested to many model building
theorists that light fermion masses could be generated from
radiative corrections\cite{earlyradm}, while those of the top and
bottom quarks   and  the tau lepton are generated at
tree level. This may be understood as the
breaking of  a symmetry among families , a horizontal symmetry .
This symmetry may be discrete \cite{modeldiscrete}, or continuous,
\cite{modelcontinuous}. The radiative generation of the light
fermions may be mediated by scalar particles as it is proposed,
for instance, in references \cite{modelrad,medscalars} and the
author in \cite{prd2007}, or also through vectorial bosons as it
happens for instance in "Dynamical Symmetry Breaking" (DSB) and
theories like " Extended Technicolor " \cite{DSB}.

\vspace{3mm}
In this report, we address the problem of fermion masses and
quark mixing within an extension of the SM introduced by the
author in \cite{albinosu32004}, which includes a vector gauged
$SU(3)$\cite{su3models} family symmetry commuting with the SM group. In previous
reports\cite{albinosu3bled} we showed that this model has the
properties to accommodate a realistic spectrum of charged fermion
masses and quark mixing. We introduce a hierarchical mass
generation mechanism in which the light fermions obtain masses
through one loop radiative corrections, mediated by the massive
bosons associated to the $SU(3)$ family symmetry that is
spontaneously broken, while the masses for the top and bottom
quarks as well as for the tau lepton, are generated at tree level
from "Dirac See-saw"\cite{SU3MKhlopov} mechanisms implemented by
the introduction of a new generation of $SU(2)_L$ weak singlets
vector-like fermions.

\vspace{3mm}
Recently, some authors have pointed out
interesting features regarding the possibility of the existence of
vector-like matter, both from theory and current
experiments\cite{vectorlikepapers}. From  the fact that the
vector-like quarks do not couple to the $W$ boson, the mixing of $U$ and $D$
vector-like quarks with the SM quarks gives rise to an extended $4\times
4$ non-unitary CKM quark mixing matrix. It has pointed out for
some authors that these vector-like fermions are weakly constrained from
Electroweak Precision Data (EWPD) because they do not break
directly the custodial symmetry, then main experimental
constraints on the vector-like matter come from the direct production
bounds,  and their implications on flavor physics. See the ref.
\cite{vectorlikepapers} for further details on
constraints for vector-like fermions. Theories and models
with extra matter may also provide attractive scenarios for present cosmological
problems, such as candidates for the nature of the Dark Matter
(\cite{normaapproach},\cite{khlopov}).

\vspace{3mm}
\emph{In this article, we report for the first time a global fit of the
free parameters of the $SU(3)$ family symmetry model to accommodate quark and
lepton masses and mixing, including light sterile neutrinos.}

\section{Model with $SU(3)$ flavor symmetry}

\subsection{Fermion content}

We define the gauge group symmetry $G\equiv SU(3) \otimes G_{SM}$
, where $SU(3)$ is a flavor symmetry among families and
$G_{SM}\equiv SU(3)_C \otimes SU(2)_L \otimes U(1)_Y$ is the
"Standard Model" gauge group, with $g_s$, $g$ and $g^\prime$ the corresponding
coupling constants. The content of fermions assumes the ordinary quarks and
leptons assigned under G as:

\begin{equation*}
\psi_q^o = ( 3 , 3 , 2 , \frac{1}{3} )_L  \qquad ,\qquad \psi_u^o = ( 3 , 3, 1 , \frac{4}{3} )_R
\qquad ,\qquad \psi_d^o = (3, 3 , 1 , -\frac{2}{3} )_R  \end{equation*}

\begin{equation*} \psi_l^o =( 3 , 1 , 2 , -1 )_L \qquad , \qquad \psi_e^o = (3, 1 , 1,-2)_R  \;, \end{equation*}

\noindent where the last entry corresponds to the
hypercharge $Y$, and the electric charge is defined by $Q = T_{3L}
+ \frac{1}{2} Y$. The model also includes two types of extra
fermions: Right handed neutrinos $\Psi_\nu^o = ( 3 , 1 , 1 , 0
)_R$, and the $SU(2)_L$ singlet vector-like fermions

 \begin{equation}
U_{L,R}^o= ( 1 , 3 , 1 , \frac{4}{3} )  \qquad , \qquad D_{L,R}^o
= ( 1 , 3 , 1 ,- \frac{2}{3} )  \label{vectorquarks} \end{equation}

\begin{equation}
N_{L,R}^o= ( 1 , 1 , 1 , 0 )
\qquad , \qquad E_{L,R}^o= ( 1 , 1 , 1 , -2 ) \label{vectorleptons}\end{equation}

The transformation of these vector-like fermions allows the mass invariant mass terms

\begin{equation}
M_U \:\bar{U}_L^o \:U_R^o \,+\, M_D \:\bar{D}_L^o \:D_R^o \,+\, M_E \:\bar{E}_L^o \:E_R^o + h.c. \;,
\end{equation}

and

\begin{equation}
m_D \,\bar{N}_L^o \,N_R^o \,+\, m_L \,\bar{N}_L^o\, (N_L^o)^c \,+\, m_R \,\bar{N}_R^o\, (N_R^o)^c \,+\,  h.c
\end{equation}

The above fermion content
make the model anomaly free. After the definition of the gauge
symmetry group and the assignment of the ordinary fermions in the
usual form under the standard model group and in the
fundamental $3$-representation under the $SU(3)$ family symmetry,
the introduction of the right-handed neutrinos is required to
cancel anomalies\cite{T.Yanagida1979}. The $SU(2)_L$ weak singlets
vector-like fermions have been introduced to give masses at tree
level only to the third family of known fermions through Dirac
See-saw mechanisms. These vector like fermions play a crucial role
to implement a hierarchical spectrum for quarks and charged lepton
masses,  together with the radiative corrections.

\section{$SU(3)$ family symmetry breaking}

To implement a hierarchical spectrum for charged fermion masses,
and simultaneously to achieve the SSB of $SU(3)$, we introduce the
flavon scalar fields: $\eta_i,\;i=2,3$, transforming under the gauge
group as $(3 , 1 , 1 , 0)$ and taking the "Vacuum Expectation
Values" (VEV's):

\begin{equation} \langle \eta_3 \rangle^T = ( 0 , 0,
\Lambda_3) \quad , \quad \langle \eta_2 \rangle^T = ( 0 ,
\Lambda_2,0)  \:. \label{veveta2eta3} \end{equation}

\noindent The above scalar fields and VEV's break completely the
$SU(3)$ flavor symmetry. The corresponding $SU(3)$ gauge bosons
are defined in Eq.(\ref{SU3lagrangian}) through their couplings to
fermions. Thus, the contribution to the horizontal gauge boson masses
from Eq.(\ref{veveta2eta3}) read

\begin{itemize}
\item $\eta_3:\quad \frac{g_{H_3}^2 \Lambda_3^2}{2} ( Y_2^+ Y_2^- + Y_3^+ Y_3^-) + g_{H_3}^2 \Lambda_3^2
\frac{Z_2^2}{3} $

\item $\eta_2:\quad \frac{g_{H_2}^2 \Lambda_2^2}{2} ( Y_1^+ Y_1^- + Y_3^+ Y_3^-) +  \frac{g_{H_2}^2 \Lambda_2^2}{4} ( Z_1^2 + \frac{Z_2^2}{3} - 2 Z_1 \frac{Z_2}{ \sqrt{3}} ) $
\end{itemize}

\noindent Therefore, neglecting tiny contributions from electroweak symmetry breaking,
we obtain the gauge boson mass terms

\begin{equation} M_1^2 \,Y_1^+ Y_1^- + \frac{M_1^2}{2} \,Z_1^2 + (\frac{4}{3} \,M_2^2 + \frac{1}{3} \,M_1^2) \frac{Z_2^2}{2} - \frac{M_1^2}{ \sqrt{3} } \,Z_1 \,Z_2  + M_2^2 \,Y_2^+ Y_2^- + ( M_1^2 +  M_2^2 ) \,Y_3^+ Y_3^-
\end{equation}

\begin{equation} M_1^2=\frac{g_{H_2}^2 \Lambda_2^2}{2} \qquad , \qquad  M_2^2= \frac{g_{H_3}^2 \Lambda_3^2}{2} \qquad , \qquad M_3^2=M_1^2 + M_2^2
\label{M1M2} \end{equation}

From the diagonalization of the $Z_1-Z_2$ squared mass matrix, we obtain
the eigenvalues

\begin{equation}
M_-^2=\frac{2}{3} \left( M_1^2 + M_2^2 - \sqrt{ (M_2^2 -  M_1^2)^2+ M_1^2 M_2^2 } \right) \quad ,\quad   M_+^2=\frac{2}{3} \left( M_1^2 + M_2^2 + \sqrt{ (M_2^2 -  M_1^2)^2+ M_1^2 M_2^2 } \right)
\label{MmMp} \end{equation}

\begin{equation} M_1^2 \,Y_1^+ Y_1^- + M_-^2 \,\frac{Z_-^2}{2} +  M_+^2 \,\frac{Z_+^2}{2}  + M_2^2 \,Y_2^+ Y_2^- + ( M_1^2 +  M_2^2 ) \,Y_3^+ Y_3^-
\end{equation}

\noindent where

\begin{equation}
\begin{pmatrix} Z_1 \\ Z_2  \end{pmatrix} = \begin{pmatrix} \cos\phi & - \sin\phi \\
\sin\phi & \cos\phi  \end{pmatrix} \begin{pmatrix} Z_- \\ Z_+  \end{pmatrix}
\end{equation}

\begin{equation*} \cos\phi \, \sin\phi=\frac{\sqrt{3}}{4} \, \frac{1}{\sqrt{(\frac{M_2^2}{M_1^2}-1)^2+ \frac{M_2^2}{M_1^2}}}
\end{equation*}

\begin{table} \begin{center} \begin{tabular}{ c | c c }
   &  $Z_1$ & $Z_2$ \\
\hline
$Z_1$ &   $M_1^2$ &  $- \frac{M_1^2}{\sqrt{3}}$ \\
      &           &                            \\
$Z_2$  & $- \frac{ M_1^2}{\sqrt{3}}$ & $\quad ( \,\frac{4}{3} \,M_2^2 + \frac{1}{3} \,M_1^2\,)$
\end{tabular} \end{center} \end{table}

\noindent with the hierarchy $M_1 , M_2 \gg M_W$.

\section{Electroweak symmetry breaking}

Recently ATLAS\cite{ATLAS} and CMS\cite{CMS} at the Large Hadron Collider announced
the discovery of a Higgs-like particle, whose properties, couplings to fermions
and gauge bosons will determine whether it is the SM Higgs or a member of an extended
Higgs sector associated to a BSM theory.  The electroweak symmetry breaking in the
$SU(3)$ family symmetry model involves the introduction of two triplets of $SU(2)_L$
Higgs doublets.

\vspace{4mm}
To achieve the spontaneous breaking of the electroweak symmetry to
$U(1)_Q$,  we introduce the scalars: $\Phi^u = ( 3 , 1 , 2 , -1 )$
and $\Phi^d = ( 3 , 1 , 2 , +1 )$, with the VEV´s: $\langle
\Phi^u \rangle^T = ( \langle \Phi_1^u \rangle , \langle \Phi_2^u \rangle
, \langle \Phi_3^u \rangle )$ , $\langle \Phi^d \rangle^T = (
\langle \Phi_1^d \rangle , \langle \Phi_2^d \rangle
, \langle \Phi_3^d \rangle )$, where $T$ means transpose,
and

\begin{equation} \qquad \langle \Phi_i^u \rangle = \frac{1}{\sqrt[]{2}} \left(
\begin{array}{c} v_i
\\ 0  \end{array} \right) \qquad , \qquad
\langle \Phi_i^d \rangle = \frac{1}{\sqrt[]{2}} \left(
\begin{array}{c} 0
\\ V_i  \end{array} \right) \:.\end{equation}

\vspace{3mm}
\noindent The contributions from
$\langle \Phi^u \rangle$ and $\langle \Phi^d \rangle$ generate
the $W$ and $Z$ gauge boson masses

\begin{equation} \frac{g^2 }{4} \,(v_u^2+v_d^2)\,
W^{+} W^{-} + \frac{ (g^2 + {g^\prime}^2) }{8}  \,(v_u^2+v_d^2)\,Z_o^2   \end{equation}

\noindent $v_u^2=v_1^2+v_2^2+v_3^2$ , $v_d^2=V_1^2+V_2^2+V_3^2 $.  Hence,
if we define as usual $M_W=\frac{1}{2} g v$, we may write $ v=\sqrt{v_u^2+v_d^2 } \thickapprox
246$ GeV.

\section{Tree level neutrino masses}

Now we describe briefly the procedure to get the masses for
ordinary fermions. The analysis for quarks and charged leptons has
already discussed in \cite{albinosu3bled}. Here, we introduce the procedure
for neutrinos.

Before "Electroweak Symmetry Breaking"(EWSB) all ordinary, "Standard Model"(SM) fermions
remain massless, and the quarks and leptons global symmetry is:

\begin{equation}
SU(3)_{q_L}\otimes SU(3)_{u_R}\otimes SU(3)_{d_R}\otimes
SU(3)_{l_L}\otimes SU(3)_{\nu_R}\otimes SU(3)_{e_R}
\end{equation}

\subsection{Tree level Dirac neutrino masses}

With the fields of particles introduced in
the model, we may write the Dirac type gauge invariant Yukawa couplings

\begin{equation} h_D \,\bar{\Psi}_l^o \,\Phi^u \,N_R^o\;\;+\;\;
h_2 \,\bar{\Psi}_\nu^o \,\eta_2 \,N_L^o \;\;+\;\;h_3 \,\bar{\Psi}_\nu^o \,\eta_3 \,N_L^o  \;\;+\;\ M_D \,\bar{N}_L^o \,N_R^o \;\;+ h.c
\label{nutlDirac} \end{equation}

\noindent $h_D$, $h_2$ and $h_3$ are Yukawa couplings, and $M_D$  a Dirac type,
invariant neutrino mass for the sterile neutrinos $N_{L,R}^o$. After electroweak
symmetry breaking, we obtain in the interaction basis ${\Psi_\nu^o}_{L,R}^T = ( \nu_e^o , \nu_\mu^o , \nu_\tau^o , N^o )_{L,R}$,  the mass terms

\begin{equation} h_D \,\left[  v_ 1 \,\bar{\nu}_{e L}^o + v_ 2 \,\bar{\nu}_{\mu L}^o + v_ 3 \, \bar{\nu}_{\tau L}^o \right] \,N_R^o + \left[  h_2\,\Lambda_2 \,\bar{\nu}_{\mu R}^o + h_3 \,\Lambda_3 \,\bar{\nu}_{\tau R}^o \right] \,N_L^o  + M_D \,\bar{N}_L^o \,N_R^o + h.c. \, ,
\end{equation}

\subsection{Tree level Majorana masses:}

 Since $N_{L,R}^o$, Eq.(\ref{vectorleptons}), are completely sterile neutrinos, we may also write the left and right handed Majorana type couplings

\begin{equation} h_L \,\bar{\Psi}_l^o \,\Phi^u (N_L^o)^c  \quad + \quad m_L \,\bar{N}_L^o\, (N_L^o)^c + h.c \end{equation}

\noindent and

\begin{equation} h_{2 R} \,\bar{\Psi}_\nu^o \,\eta_2 \,(N_R^o)^c \;\;+\;\;h_{3 R} \,\bar{\Psi}_\nu^o \,\eta_3 \,(N_R^o)^c \;\;+\;\;m_R \,\bar{N}_R^o \,(N_R^o)^c + h.c \; ,\end{equation}

\noindent respectively. After spontaneous symmetry breaking, we also get the left handed and right handed Majorana mass terms

\begin{equation} h_L \,\left[  v_ 1 \,\bar{\nu}_{e L}^o + v_ 2 \,\bar{\nu}_{\mu L}^o + v_ 3 \, \bar{\nu}_{\tau L}^o         \right] \,(N_L^o)^c  \quad + \quad m_L \,\bar{N}_L^o \,(N_L^o)^c  + h.c. \, ,
\label{nutlML} \end{equation}

\begin{equation}
+ \left[  h_{2 R} \,\Lambda_2 \,\bar{\nu}_{\mu R}^o + h_{3 R} \,\Lambda_3 \,\bar{\nu}_{\tau R}^o  \right] \,(N_R^o)^c  \;\;+\;\;m_R \,\bar{N}_R^o \,(N_R^o)^c + h.c. \, ,
\label{nutlMR} \end{equation}

\begin{table}[h]
\begin{center}
\begin{tabular}{ c | c c c c c c c c}
&$(\nu^o_{e L})^c$ & $(\nu^o_{\mu L})^c$ & $(\nu^o_{\tau L})^c$ & $\nu^o_{e R}$ & $\nu^o_{\mu R}$ & $\nu^o_{\tau R}$ & $(N^o_L)^c$ & $N_R^o$ \\
\hline
$\overline{\nu^o_{e L}} $  & 0 & 0 & 0 & 0 &  0 & 0 & $h_L v_1$ & $h_D v_1$  \\
                                                                        \\
$\overline{\nu^o_{\mu L} }$  & 0 & 0 & 0 & 0 &  0 & 0 & $h_L v_2$  & $h_D  v_2$  \\
                                                                          \\
$\overline{\nu^o_{\tau L} }$  & 0 & 0 & 0 & 0 & 0 & 0 & $h_L v_3$ & $h_D  v_3$  \\
                                                                           \\
$\overline{(\nu^o_{e R})^c}$  & 0 & 0 & 0 & 0 &  0 & 0 & 0 & 0  \\
                                                                \\
$\overline{(\nu^o_{\mu R})^c}$  & 0 & 0 & 0 & 0 &  0 & 0 & $h_2 \Lambda_2$ & $h_{2 R} \Lambda_2$  \\
                                                                \\
$\overline{(\nu^o_{\tau R})^c}$  & 0 & 0 & 0 & 0 &  0 & 0 & $h_3 \Lambda_3$ & $h_{3 R} \Lambda_3$  \\
                                                                \\
$\overline{N^o_L}$  & $h_L v_1$  & $h_L v_2$ & $h_L v_3$ & 0 & $h_2  \Lambda_2$ & $h_3 \Lambda_3$ & $m_L$  & $M_D$  \\
                                                              \\
$\overline{(N^o_R)^c}$ & $h_D  v_1$  & $h_D  v_2$  & $h_D  v_3$  & 0 & $h_{2 R} \Lambda_2$  & $h_{3 R} \Lambda_3$ & $M_D$  & $m_R$
\end{tabular} \end{center}
\caption{Tree Level Majorana masses }
\label{treelevelMajorana} \end{table}

\vspace{3mm}
Thus, in the basis ${\Psi_\nu^o}^T= \left( \,\nu^o_{e L} \, , \, \nu^o_{\mu L} \, , \,  \nu^o_{\tau L} \, , \, (\nu^o_{e R})^c \, , \, (\nu^o_{\mu R})^c \, , \, (\nu^o_{\tau R})^c \, , \, N^o_L   \, , \, (N^o_R)^c \, \right)$, the Generic $8\times 8$ tree level Majorana mass matrix for neutrinos $\mathcal{M}_\nu^o$, from Table \ref{treelevelMajorana}, $\bar{\Psi_\nu^o}\; \mathcal{M}_\nu^o \;(\Psi_\nu^o)^c \;+h.c.$, read

\begin{equation} \mathcal{M}_\nu^o=
\begin{pmatrix}
0 & 0 & 0 & 0 &  0 & 0 & \alpha_1 & a_1  \\
0 & 0 & 0 & 0 &  0 & 0 & \alpha_2 & a_2 \\
0 & 0 & 0 & 0 & 0 & 0 & \alpha_3 & a_3 \\
0 & 0 & 0 & 0 &  0 & 0 & 0 & 0  \\
0 & 0 & 0 & 0 &  0 & 0 & b_2 & \beta_2 \\
0 & 0 & 0 & 0 &  0 & 0 & b_3 & \beta_3  \\
\alpha_1  & \alpha_2 & \alpha_3 & 0 & b_2 & b_3  & m_L  & m_D  \\
a_1  & a_2  & a_3  & 0 & \beta_2  & \beta_3 & m_D  & m_R
\end{pmatrix} \label{nuoMajorana}
\end{equation}

\vspace{4mm}
Diagonalization of $\mathcal{M}_\nu^{(o)}$, Eq.(\ref{nuoMajorana}), yields four zero eigenvalues,
associated to the neutrino fields: $ap=\sqrt{a_1^2+a_2^2}$

\begin{equation*}
\frac{a_2}{ap}\,\nu^o_{e L}-\frac{a_1}{ap}\,\nu^o_{\mu L} \quad ,\quad
\frac{a_1 \,a_3}{ap \,a}\,\nu^o_{e L}+\frac{a_2 \,a_3}{ap \,a}\,\nu^o_{\mu L}-\frac{a_p}{a}\,\nu^o_{\tau L},
\end{equation*}

\begin{equation*}
\nu^o_{e R} \quad ,\quad \frac{b_3}{b}\,\nu^o_{\mu R}-\frac{b_2}{b}\,\nu^o_{\tau R}
\end{equation*}

\vspace{3mm}
Assuming for simplicity,  $\frac{h_2}{h_{2R}}=\frac{h_3}{h_{3R}}$, the Characteristic Polynomial for the
nonzero eigenvalues of $\mathcal{M}_\nu^o$ reduce to the one of the matrix $m_4$, Eq.(\ref{m4}),
where

\vspace{3mm}
\begin{equation}  m_4=
\begin{pmatrix}
0 & 0 & \alpha & a  \\
0 & 0 & b & \beta  \\
\alpha & b  & m_L  & m_D  \\
a & \beta & m_D  & m_R
\end{pmatrix} \quad , \quad  U_4=
\begin{pmatrix}
u_{11} & u_{12} & u_{13} & u_{14} \\
u_{21} & u_{22} & u_{23} & u_{24} \\
u_{31} & u_{32} & u_{33} & u_{34} \\
u_{41} & u_{42} & u_{43} & u_{44}
\end{pmatrix} \label{m4} \end{equation}

\begin{eqnarray}  a=\sqrt{a_1^2+a_2^2+a_3^2} ,& \alpha=\sqrt{\alpha_1^2+\alpha_2^2+\alpha_3^2}\;,\nonumber \\ b=\sqrt{b_2^2+b_3^2}\;,& \beta=\sqrt{\beta_2^2+\beta_3^2} \nonumber
\end{eqnarray}

\vspace{3mm}
\begin{equation} U_4^T\, m_4\, U_4 = Diag(m_5^o, m_6^o, m_7^o, m_8^o)\equiv d_4  \quad , \quad
m_4= U_4\,d_4\,U_4^T
\label{U4} \end{equation}

Eq.(\ref{U4}) impose the constrains

\begin{eqnarray}
u_{11}^2\,m_5^o+u_{12}^2\,m_6^o+u_{13}^2\,m_7^o+u_{14}^2\,m_8^o&=&0\\ \nonumber\\
u_{21}^2\,m_5^o+u_{22}^2\,m_6^o+u_{23}^2\,m_7^o+u_{24}^2\,m_8^o&=&0\\ \nonumber\\
u_{11} u_{21} \,m_5^o+u_{12} u_{22}\,m_6^o+u_{13} u_{23}\,m_7^o+u_{14} u_{24}\,m_8^o&=&0\,,
\end{eqnarray}

\noindent corresponding to the $(m_4)_{11}=(m_4)_{22}=(m_4)_{12}=0$ zero entries, respectively.

\vspace{5mm}
In this form, we diagonalize $\mathcal{M}_\nu^o$ by using the orthogonal matrix

\vspace{3mm}
\begin{equation}  U_\nu^o=
\begin{pmatrix}
 \frac{a_2}{ap} & \frac{a_1 \,a_3}{a \,ap} & 0 & 0 & \frac{a_1}{a} u_{11} & \frac{a_1}{a} u_{12} & \frac{a_1}{a} u_{13} & \frac{a_1}{a} u_{14} \\\\
- \frac{a_1}{ap} & \frac{a_2 \,a_3}{a \,ap} &  0 & 0 & \frac{a_2}{a} u_{11} & \frac{a_2}{a} u_{12} & \frac{a_2}{a} u_{13} & \frac{a_2}{a} u_{14} \\\\
0 & -\frac{ap}{a}  & 0 & 0 & \frac{a_3}{a} u_{11} & \frac{a_3}{a} u_{12} & \frac{a_3}{a} u_{13} & \frac{a_3}{a} u_{14} \\\\
 0 & 0 & 1 & 0 & 0 & 0 & 0 & 0 \\\\
 0 & 0 & 0 & \frac{b_3}{b} & \frac{b_2}{b} u_{21} & \frac{b_2}{b} u_{22} & \frac{b_2}{b} u_{23} & \frac{b_2}{b} u_{24} \\\\
0 & 0 & 0 & -\frac{b_2}{b} & \frac{b_3}{b} u_{21} & \frac{b_3}{b} u_{22} & \frac{b_3}{b} u_{23} & \frac{b_3}{b} u_{24} \\\\
 0 & 0 & 0 & 0 & u_{31} & u_{32} & u_{33} & u_{34} \\\\
 0 & 0 & 0 & 0 & u_{41} & u_{42} & u_{43} & u_{44}
\end{pmatrix} \end{equation}

\vspace{5mm}
\begin{equation} (U_\nu^o)^T\,\mathcal{M}_\nu^o \,U_\nu^o   = Diag(0,0,0,0, m_5^o, m_6^o, m_7^o, m_8^o) \end{equation}

\vspace{3mm}
\emph{Notice that the first four columns in $U_\nu^o$ correspond to the four massless eigenvectors.
Hence, the tree level mixing, $U_\nu^o$, depends on the ordering we define for these four degenerated
massless eigenvectors. However, it turns out that the final mixing product $U_\nu^o\;U_\nu$, as well as
the final mass eigenvalues are independent of the choice of this ordering.}

\section{One loop neutrino masses}

After tree level contributions the fermion global symmetry is broken down to:

\begin{equation}
SU(2)_{q_L}\otimes SU(2)_{u_R}\otimes SU(2)_{d_R}\otimes
SU(2)_{l_L}\otimes SU(2)_{\nu_R}\otimes SU(2)_{e_R}
\end{equation}

Therefore, in this scenario light neutrinos may get extremely small masses from
radiative corrections mediated by the $SU(3)$ heavy gauge bosons.

\vspace{3mm}
\subsection{One loop Dirac Neutrino masses}

After the breakdown of the electroweak symmetry, neutrinos may get
tiny Dirac mass terms from the generic one loop diagram in Fig. 1, The internal fermion line
in this diagram represent the tree level see-saw mechanisms, Eqs.(\ref{nutlDirac}-\ref{nutlMR}). The vertices
read from the $SU(3)$ family symmetry interaction Lagrangian

\begin{multline} i {\cal{L}}_{int} = \frac{g_{H}}{2}
\left( \bar{e^{o}}
\gamma_{\mu} e^{o}- \bar{\mu^{o}} \gamma_{\mu} \mu^{o} \right) Z_1^\mu
+  \frac{g_{H}}{2 \sqrt{3}} \left( \bar{e^{o}} \gamma_{\mu} e^{o}+ \bar{\mu^{o}}
\gamma_{\mu} \mu^{o} - 2 \bar{\tau^{o}}
\gamma_{\mu} \tau^{o}  \right) Z_2^\mu                \\
+ \frac{g_{H}}{\sqrt{2}} \left( \bar{e^{o}} \gamma_{\mu} \mu^{o} Y_1^{+} +
\bar{e^{o}} \gamma_{\mu} \tau^{o} Y_2^{+} + \bar{\mu^{o}} \gamma_{\mu} \tau^{o} Y_3^{+} + h.c.
\right) \label{SU3lagrangian} \end{multline}

\noindent The contribution from these diagrams may be written as

\begin{equation} c_Y \frac{\alpha_H}{\pi}\,m_\nu(M_Y)_{ij} \quad ,\quad \alpha_H = \frac{g_H^2}{4 \pi} \, , \end{equation}

\noindent
\begin{equation}m_\nu(M_Y)_{ij} \equiv \sum_{k=5,6,7,8} m_k^o \:U^o_{ik} U^o_{jk}\:f(M_Y, m_k^o) \end{equation}

\noindent and $f_{Y_k}=\frac{M_Y^2}{M_Y^2 - m_k^{o\,2}} \,ln{\frac{M_Y^2}{m_k^{o\,2}}} \approx ln{\frac{M_Y^2}{m_k^{o\,2}}}$

\begin{figure}[h]
\centering
\includegraphics{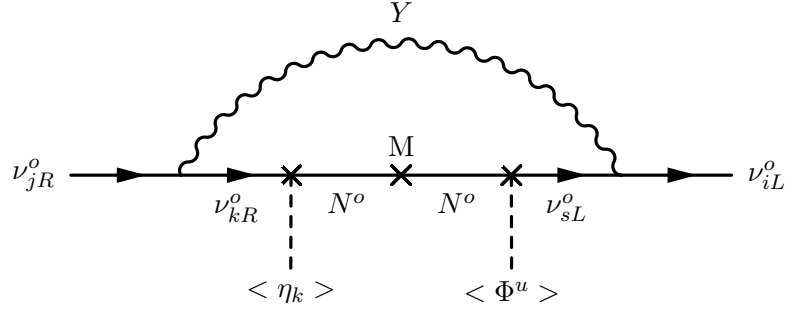}
\caption{ Generic one loop diagram contribution to the Dirac mass term
$m_{ij} \:{\bar{\nu}}_{iL}^o \nu_{jR}^o$. $\; \text{M}=M_D , m_L, m_R$}
\end{figure}

\begin{table}[h]
\begin{center}
\begin{tabular}{ c | c c c c }
   &  $\nu^o_{e R}$ & $\nu^o_{\mu R} $ & $\nu^o_{\tau R} $ & $N^o_R $ \\
\hline
$\bar{\nu}^o_{e L}$  & $D_{\nu \,14}$  & $D_{\nu \,15}$  & $D_{\nu \,16}$   & 0  \\
                                                                        \\
$\bar{\nu}^o_{\mu L}$  & 0 & $D_{\nu \,25}$  & $D_{\nu \,26}$    & 0  \\
                                                                          \\
$\bar{\nu}^o_{\tau L}$  & 0  &  $D_{\nu \,35}$  & $D_{\nu \,36}$ & 0   \\
                                                                           \\
$\bar{N}^o_L$  &  0 & 0 & 0 & 0
\end{tabular} \end{center}
\caption{One loop Dirac mass terms $m_{ij} \:{\bar{\nu}}_{iL}^o \;\nu_{jR}^o$ }
\end{table}

\begin{align*}
m_\nu(M_{Z_1})_{ij}=\, & \cos\phi \,m_\nu(M_-)_{ij} - \sin\phi \,m_\nu(M_+)_{ij}     \\\\
m_\nu(M_{Z_2})_{ij}=\, & \sin\phi \,m_\nu(M_-)_{ij} + \cos\phi \,m_\nu(M_+)_{ij}       \\\\
G_{\nu,m \,ij}=\, &  \frac{\sqrt{\alpha_2 \alpha_3} }{\pi} \,\frac{1}{2\sqrt{3}}
\,\cos\phi \,\sin\phi \,[m_\nu(M_-)_{ij} - m_\nu(M_+)_{ij}  ]
\end{align*}

\begin{align}
\mathcal{F}(M_Y)=\, & m_5^o \,u_{11}^2 \,f_{Y_5} + m_6^o \,u_{12}^2 \,f_{Y_6} + m_7^o \,u_{13}^2 \,f_{Y_7} + m_8^o \,u_{14}^2 \,f_{Y_8}      \\  \nonumber\\
\mathcal{G}(M_Y)=\, & m_5^o \,u_{21}^2 \,f_{Y_5} + m_6^o \,u_{22}^2 \,f_{Y_6} + m_7^o \,u_{23}^2 \,f_{Y_7} + m_8^o \,u_{24}^2 \,f_{Y_8}
\end{align}

\begin{equation}
\mathcal{H}(M_Y)=\, m_5^o \,u_{11} u_{21} \,f_{Y_5} + m_6^o \,u_{12} u_{22} \,f_{Y_6} + m_7^o \,u_{13} u_{23} \,f_{Y_7} + m_8^o \,u_{14} u_{24} \,f_{Y_8}
\end{equation}

\begin{eqnarray*}
m_\nu(M_Y)_{15} = \frac{a_1 b_2}{a b} \mathcal{H}(M_Y) &;\quad m_\nu(M_Y)_{16} = \frac{a_1 b_3}{a b} \mathcal{H}(M_Y)  \\\\
m_\nu(M_Y)_{25} = \frac{a_2 b_2}{a b} \mathcal{H}(M_Y) &;\quad m_\nu(M_Y)_{26} = \frac{a_2 b_3}{a b} \mathcal{H}(M_Y)  \\\\
m_\nu(M_Y)_{35} = \frac{a_3 b_2}{a b} \mathcal{H}(M_Y) &;\quad m_\nu(M_Y)_{36} = \frac{a_3 b_3}{a b} \mathcal{H}(M_Y)
\end{eqnarray*}

\begin{align*}
D_{\nu \,14} = & \; \frac{1}{2} \left[ \frac{a_2 b_2}{a b} \mathcal{H}(M_1) + \frac{a_3 b_3}{a b} \mathcal{H}(M_2)       \right]  \; , \\\\
D_{\nu \,15} = & \; \frac{a_1 b_2}{a b} \,\left[ -\frac{1}{4} \mathcal{H}(M_{Z_1}) + \frac{1}{12} \mathcal{H}(M_{Z_2})       \right]    \; , \\\\
D_{\nu \,25} = & \;\frac{a_2 b_2}{a b} \,\left[ \frac{1}{4} \mathcal{H}(M_{Z_1}) + \frac{1}{12} \mathcal{H}(M_{Z_2}) - \mathcal{H}(G_{\nu , m}) \right]  +  \frac{1}{2} \frac{a_3 b_3}{a b} \mathcal{H}(M_3)       \; ,\\\\
D_{\nu \,36} = & \;\frac{1}{2} \frac{a_2 b_2}{a b} \mathcal{H}(M_3) + \frac{1}{3} \frac{a_3 b_3}{a b} \mathcal{H}(M_{Z_2})       \; ,\\\\
D_{\nu \,16} = & \;\frac{a_1 b_3}{a b} \left[  - \frac{1}{6} \mathcal{H}(M_{Z_2})  - \mathcal{H}(G_{\nu , m})  \right]   \; , \\\\
D_{\nu \,26} = & \;\frac{a_2 b_3}{a b} \left[  - \frac{1}{6} \mathcal{H}(M_{Z_2})  + \mathcal{H}(G_{\nu , m})  \right]    \; ,\\\\
D_{\nu \,35} = & \;\frac{a_3 b_2}{a b} \left[  - \frac{1}{6} \mathcal{H}(M_{Z_2})  + \mathcal{H}(G_{\nu , m})  \right]    \; ,
\end{align*}

\begin{equation*}
\mathcal{H}(G_{\nu , m}) =  \frac{\sqrt{\alpha_2 \alpha_3} }{\pi} \,\frac{1}{2\sqrt{3}}
\,\cos\phi \,\sin\phi \,[\mathcal{H}(M_- ) - \mathcal{H}(M_+ )]
\end{equation*}

\vspace{5mm}
\subsection{One loop L-handed Majorana masses}

Neutrinos also obtain one loop corrections to L-handed and R-handed Majorana masses from
the diagrams of Fig. 2 and Fig. 3, respectively.

\begin{figure}[h] \centering
\includegraphics{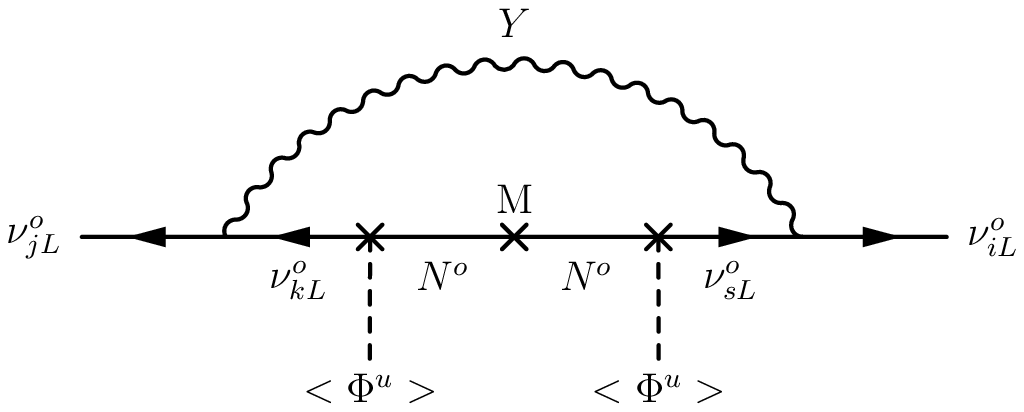}
\caption{ Generic one loop diagram contribution to the L-handed Majorana mass term
$m_{ij} \:{\bar{\nu}}_{iL}^o (\nu_{jL}^o)^T$. $\; \text{M}=M_D , m_L, m_R$}
\end{figure}

A similar procedure as for Dirac Neutrino masses,
leads to the one loop Majorana mass terms

\vspace{4mm}
\begin{table}[h]
\begin{center}
\begin{tabular}{ c | c c c c }
   &  $\nu^o_{e L}$ & $\nu^o_{\mu L} $ & $\nu^o_{\tau L} $ & $N^o_L $ \\
\hline
$\nu^o_{e L}$  &  $L_{\nu \,11}$ & $L_{\nu \,12}$ &  $L_{\nu \,13}$   & 0  \\
                                                                        \\
$\nu^o_{\mu L}$  &  $L_{\nu \,12}$ & $L_{\nu \,22}$  & $L_{\nu \,23}$   & 0  \\
                                                                          \\
$\nu^o_{\tau L}$  & $L_{\nu \,13}$  & $L_{\nu \,23}$  & $L_{\nu \,33}$   & 0   \\
                                                                           \\
$N^o_L$  &  0 &\quad 0 &\quad 0 & 0
\end{tabular} \end{center}
\caption{One loop L-handed Majorana mass terms $m_{ij} \:{\bar{\nu}}_{iL}^o \;(\nu_{jL}^o)^T$ }
\end{table}

\begin{eqnarray*}
m_\nu(M_Y)_{11} = \frac{a_1^2}{a^2} \mathcal{F}(M_Y) &;\quad m_\nu(M_Y)_{12} = \frac{a_1 a_2}{a^2} \mathcal{F}(M_Y)
\\\\
m_\nu(M_Y)_{13} = \frac{a_1 a_3}{a^2} \mathcal{F}(M_Y) &;\quad m_\nu(M_Y)_{22} = \frac{a_2^2}{a^2} \mathcal{F}(M_Y)
\\\\
m_\nu(M_Y)_{23} = \frac{a_2 a_3}{a^2} \mathcal{F}(M_Y) &;\quad m_\nu(M_Y)_{33} = \frac{a_3^2}{a^2} \mathcal{F}(M_Y)
\end{eqnarray*}

\begin{align*}
L_{\nu \,11} = & \; \frac{a_1^2}{a^2} \left[ \frac{1}{4} \mathcal{F}(M_{Z_1}) + \frac{1}{12} \mathcal{F}(M_{z_2})       + \mathcal{F}(G_{\nu , m}) \right]  \; , \\\\
L_{\nu \,22} = & \; \frac{a_2^2}{a^2} \left[ \frac{1}{4} \mathcal{F}(M_{Z_1}) + \frac{1}{12} \mathcal{F}(M_{z_2})       - \mathcal{F}(G_{\nu , m}) \right]  \; , \\\\
L_{\nu \,33} = & \;\frac{1}{3} \frac{a_3^2}{a^2}  \mathcal{F}(M_{z_2})  \; , \\\\
L_{\nu \,12} = & \; \frac{a_1 a_2}{a^2} \left[ -\frac{1}{4} \mathcal{F}(M_{Z_1}) + \frac{1}{2} \mathcal{F}(M_1)  + \frac{1}{12} \mathcal{F}(M_{z_2}) \right]  \; , \\\\
L_{\nu \,13} = & \; \frac{a_1 a_3}{a^2} \left[ -\frac{1}{6} \mathcal{F}(M_{Z_2}) + \frac{1}{2} \mathcal{F}(M_2)  - \mathcal{F}(G_{\nu , m}) \right]  \; , \\\\
L_{\nu \,23} = & \; \frac{a_2 a_3}{a^2} \left[ -\frac{1}{6} \mathcal{F}(M_{Z_2}) + \frac{1}{2} \mathcal{F}(M_3)  + \mathcal{F}(G_{\nu , m}) \right]
\end{align*}

\begin{align}
\mathcal{F}(G_{\nu , m}) &=  \frac{\sqrt{\alpha_2 \alpha_3} }{\pi} \,\frac{1}{2\sqrt{3}}
\,\cos\phi \,\sin\phi \,[\mathcal{F}(M_- ) - \mathcal{F}(M_+ )]
\end{align}

\vspace{5mm}
\subsection{One loop R-handed Majorana masses}

\begin{figure}[h] \centering
\includegraphics{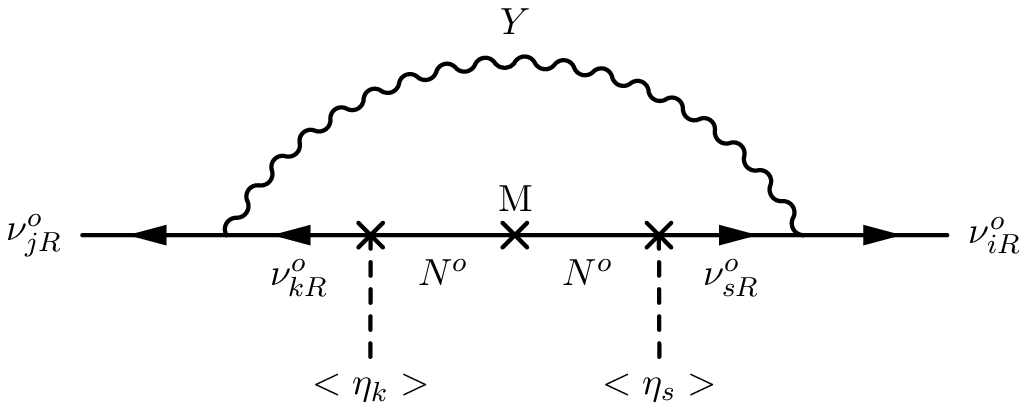}
\caption{ Generic one loop diagram contribution to the R-handed Majorana mass term
$m_{ij} \:{\bar{\nu}}_{iR}^o (\nu_{jR}^o)^T$. $\; \text{M}=M_D , m_L, m_R$}
\end{figure}

\vspace{5mm}
\begin{table}[h]
\begin{center}
\begin{tabular}{ c | c c c c }
   &  $\nu^o_{e R}$ & $\nu^o_{\mu R} $ & $\nu^o_{\tau R} $ & $N^o_R $ \\
\hline
$\nu^o_{e R}$  & 0  &\quad 0  &\quad 0 & 0  \\
                                                                        \\
$\nu^o_{\mu R}$  & 0 & $R_{\nu \,55}$  & $R_{\nu \,56}$  & 0  \\
                                                                          \\
$\nu^o_{\tau R}$  & 0 & $R_{\nu \,56}$  & $R_{\nu \,66}$   & 0   \\
                                                                 \\
$N^o_R$  &  0 &\quad 0 &\quad 0 & 0
\end{tabular} \end{center}
\caption{One loop R-handed Majorana mass terms $m_{ij} \:{\bar{\nu}}_{iR}^o \;(\nu_{jR}^o)^T$ }
\end{table}
\vspace{5mm}

\begin{eqnarray*}
m_\nu(M_Y)_{55} = \frac{b_2^2}{b^2} \mathcal{G}(M_Y) &;\quad m_\nu(M_Y)_{66} = \frac{b_3^2}{b^2} \mathcal{G}(M_Y) \\\\
m_\nu(M_Y)_{56} = \frac{b_2 b_3}{b^2} \mathcal{G}(M_Y) \nonumber  \\
\end{eqnarray*}

\begin{align*}
R_{\nu \,55} = & \; \frac{b_2^2}{b^2} \left[ \frac{1}{4} \mathcal{G}(M_{Z_1}) + \frac{1}{12} \mathcal{G}(M_{z_2})       - \mathcal{G}(G_{\nu , m}) \right]  \; , \\\\
R_{\nu \,66} = & \;\frac{1}{3} \frac{b_3^2}{b^2}  \mathcal{G}(M_{z_2})  \; , \\\\
R_{\nu \,56} = & \; \frac{b_2 b_3}{b^2} \left[ -\frac{1}{6} \mathcal{G}(M_{Z_2}) + \frac{1}{2} \mathcal{G}(M_3)  + \mathcal{G}(G_{\nu , m}) \right]
\end{align*}

\begin{align*}
\mathcal{G}(G_{\nu , m}) &=  \frac{\sqrt{\alpha_2 \alpha_3} }{\pi} \,\frac{1}{2\sqrt{3}}
\,\cos\phi \,\sin\phi \,[\mathcal{G}(M_- ) - \mathcal{G}(M_+ )]
\end{align*}

Thus, in the $\Psi_\nu^o$ basis,  we may write the one loop contribution for neutrinos as

\vspace{3mm}
\begin{equation} \mathcal{M}_{1\,\nu}^o=
\begin{pmatrix}
L_{\nu \,11} &\quad L_{\nu \,12} &\quad L_{\nu \,13} &\quad D_{\nu \,14}  &\quad D_{\nu \,15}  &\quad D_{\nu \,16}   &\quad 0 &\quad 0  \\
                                                                        \\
L_{\nu \,12} &\quad L_{\nu \,22}  &\quad L_{\nu \,23} &\quad 0 &\quad D_{\nu \,25}  &\quad D_{\nu \,26}  &\quad 0 &\quad 0   \\
                                                                          \\
L_{\nu \,13}  &\quad L_{\nu \,23} &\quad L_{\nu \,33}  &\quad 0 &\quad D_{\nu \,35}  &\quad D_{\nu \,36}   &\quad 0 &\quad 0 \\
                                                                           \\
D_{\nu \,14}  &\quad 0 &\quad 0 &\quad 0 &\quad 0 &\quad 0 &\quad 0 &\quad 0  \\
                                                                \\
D_{\nu \,15} &\quad D_{\nu \,25} &\quad D_{\nu \,35} &\quad 0 &\quad R_{\nu \,55} &\quad R_{\nu \,56} &\quad 0 &\quad 0  \\
                                                                \\
D_{\nu \,16} &\quad D_{\nu \,26} &\quad D_{\nu \,36} &\quad 0 &\quad R_{\nu \,56} &\quad R_{\nu \,66} &\quad 0 &\quad 0  \\
                                                                \\
0 &\quad 0 &\quad 0 &\quad 0 &\quad 0 &\quad 0 &\quad 0  &\quad 0  \\
                                                              \\
0 &\quad 0 &\quad 0 &\quad 0 &\quad 0 &\quad 0 &\quad 0 &\quad 0
\end{pmatrix} \end{equation}

\vspace{4mm}
\subsection{Neutrino mass matrix up to one loop}

Finally, we obtain the Majorana mass matrix for neutrinos up to one loop

\begin{equation} \mathcal{M}_{\nu}= ( U_\nu^o )^T\, \mathcal{M}_{1\,\nu}^o \,U_\nu^o + Diag(0, 0, 0, 0, m_5^o, m_6^o, m_7^o, m_8^o) \,, \end{equation}

\noindent where explicitly

\begin{equation} \mathcal{M}_{\nu}=
\begin{pmatrix}
N_{11} & N_{12} & N_{13} & N_{14} & N_{15} & N_{16} & N_{17} & N_{18} \\\\
N_{12} & N_{22} & N_{23} & N_{24} & N_{25} & N_{26} & N_{27} & N_{28} \\\\
N_{13} & N_{23} & 0 & 0 & N_{35} & N_{36} & N_{37} & N_{38} \\\\
N_{14} & N_{24} & 0 & N_{44} & N_{45} & N_{46} & N_{47} & N_{48} \\\\
N_{15} & N_{25} & N_{35} & N_{45} & N_{55}+m_5^o & N_{56} & N_{57} & N_{58} \\\\
N_{16} & N_{26} & N_{36} & N_{46} & N_{56} & N_{66}+m_6^o & N_{67} & N_{68} \\\\
N_{17} & N_{27} & N_{37} & N_{47} & N_{57} & N_{67} & N_{77}+m_7^o & N_{78} \\\\
N_{18} & N_{28} & N_{38} & N_{48} & N_{58} & N_{68} & N_{78} & N_{88}+m_8^o
\end{pmatrix} \end{equation}

\vspace{3mm}
\noindent {\bf Majorana L-handed:}

\begin{equation} N_{11}= \frac{a_1^2 a_2^2}{a_p^2 a^2} ( F_{Z_1} - F_1 ) \end{equation}

\begin{equation}
N_{12}= - \frac{a_1 a_2 a_3}{2 a^3} [ \frac{a_2^2 - a_1^2}{a_p^2} (F_{Z_1} - F_1) + F_2 - F_3  - 6 F_m  ]
\end{equation}

\begin{equation}
N_{22}=\frac{a_3^2}{a^2} \,\left[\, \frac{1}{4}\frac{(a_2^2 - a_1^2)^2}{a_p^2 \,a^2} (F_{Z_1} - F_1) + \frac{a_2^2}{a^2}(F_2 - F_3)  + \frac{a_p^2}{4 \,a^2}(F_1+3 F_{Z_2}  - 4 F_2) - 3 \frac{a_2^2 - a_1^2}{a^2} F_m \,\right]
\end{equation}

\vspace{3mm}
\noindent {\bf Dirac:}

\begin{equation} N_{13}=\frac{a_2}{2\,a_p} \,( \frac{a_2 \,b_2}{a\,b} H_1 + \frac{a_3 \,b_3}{a\,b} H_2 )=q_{11}
\end{equation}

\begin{equation} N_{14}=- \frac{a_1\,b_3}{2\,a_p\,b} \,( \frac{a_2 \,b_2}{a\,b} H_{Z_1} + \frac{a_3 \,b_3}{a\,b} H_3 - 6 \frac{a_2 \,b_2}{a\,b} H_m  )=q_{12}
\end{equation}

\begin{equation} N_{23}=\frac{a_1\,a_3}{2\,a_p\,a} \,( \frac{a_2 \,b_2}{a\,b} H_1 + \frac{a_3 \,b_3}{a\,b} H_2 )=q_{21} \end{equation}

\begin{equation}
N_{24}= \frac{a_2 \,(a_p^2\,b_2^2 + a_3^2\,b_3^2)}{2\,a_p\,a^2\,b^2} H_3 +
\frac{(a_2^2 - a_1^2) a_3\,b_2\,b_3}{4\,a_p\,a^2\,b^2} H_{Z_1}  +
\frac{3\,a_p\,a_3\,b_2\,b_3}{4\,a^2\,b^2} H_{Z_2} -
\frac{3\,a_2^2\,a_3\,b_2\,b_3}{a_p\,a^2\,b^2} H_m=q_{22}
\end{equation}

\vspace{3mm}
\noindent {\bf Majorana R-handed:}

\begin{equation} N_{44}=\frac{b_2^2\,b_3^2}{4\,b^4} \,(\,G_{Z_1} + 3 G_{Z_2} - 4 G_3 - 12 G_m\,)
\end{equation}

\vspace{3mm}
\noindent {\bf Majorana L-handed and Dirac:}

\begin{eqnarray}
N_{15}=- F_{15}\,u_{11} + q_{13}\,u_{21} \qquad &; \qquad N_{16}=- F_{15}\,u_{12} + q_{13}\,u_{22}  \\
\nonumber \\
N_{17}=- F_{15}\,u_{13} + q_{13}\,u_{23} \qquad &; \qquad N_{18}=- F_{15}\,u_{14} + q_{13}\,u_{24}
\end{eqnarray}

\begin{equation*}
F_{15}=\frac{a_1\,a_2}{2\,a_p\,a} \,\left[\,\frac{a_2^2 - a_1^2}{a^2} (F_{Z_1} - F_1) + \frac{a_3^2}{a^2}(F_3 - F_2) + 2\, \frac{(2\,a_3^2 - a_p^2)}{a^2} \,F_m \,\right]
\end{equation*}

\begin{equation*}
q_{13}= - \frac{a_1\,b_2}{2 a_p\,b} \,\left[ \frac{a_2 \,b_2}{a\,b} H_{Z_1} + \frac{a_3 \,b_3}{a\,b} H_3
- 2 \frac{a_2}{a}\, \frac{b_2^2 - 2 b_3^2}{b_2\,b} \,H_m  \right]
\end{equation*}

\begin{eqnarray}
N_{25}=F_{25}\,u_{11} + q_{23}\,u_{21} \qquad &; \qquad N_{26}=F_{25}\,u_{12} + q_{23}\,u_{22}  \\
\nonumber \\
N_{27}=F_{25}\,u_{13} + q_{23}\,u_{23} \qquad &; \qquad N_{28}=F_{25}\,u_{14} + q_{23}\,u_{24}
\end{eqnarray}

\begin{multline*} F_{25}=
\frac{a_3}{4\,a_p\,a^4} \,\left[ \,(a_2^2 - a_1^2)^2\,( F_{Z_1} - F_1) + 2\,a_2^2(a_3^2-a_p^2)\,(F_3 - F_2) - a_p^4\,(F_{Z_2} - F_1) \right. \\ \left.  - 2\,a_p^2(a_3^2-a_p^2)\,(F_{Z_2} - F_2) +
4\,(a_2^2 - a_1^2)\,(a_3^2 - 2 a_p^2) \,F_m \, \right]
\end{multline*}

\begin{equation*} q_{23}= \frac{a_2 \,(a_3^2 - a_p^2\,)b_2\,b_3}{2\,a_p\,a^2\,b^2} H_3 +
\frac{(a_2^2 - a_1^2) a_3\,b_2^2}{4\,a_p\,a^2\,b^2} H_{Z_1} +
\frac{a_p\,a_3\,(b_2^2 -2\,b_3^2)}{4\,a^2\,b^2} H_{Z_2} -
\frac{a_3\,[a_p^2 \,b^2+a_2^2(b_2^2 -2\,b_3^2)] }{a_p\,a^2\,b^2} H_m
\end{equation*}

\vspace{3mm}
\noindent {\bf Dirac:}

\begin{equation}
N_{35}=q_{31}\,u_{11} \; , \; N_{36}=q_{31}\,u_{12} \; , \;
N_{37}=q_{31}\,u_{13} \; , \; N_{38}=q_{31}\,u_{14}
\end{equation}

\begin{equation*} q_{31}=\frac{a_1}{2 a}\,( \frac{a_2 \,b_2}{a\,b} H_1 + \frac{a_3 \,b_3}{a\,b} H_2 )
\end{equation*}

\vspace{3mm}
\noindent {\bf Dirac and Majorana R-handed:}

\begin{eqnarray}
N_{45}=q_{32}\,u_{11} + G_{45}\,u_{21} \quad &, \quad N_{46}=q_{32}\,u_{12} + G_{45}\,u_{22}  \\
\nonumber \\
N_{47}=q_{32}\,u_{13} + G_{45}\,u_{23} \quad &, \quad N_{48}=q_{32}\,u_{14} + G_{45}\,u_{24}
\end{eqnarray}

\begin{equation*}
q_{32}= - \frac{a_2\,a_3\,(b_2^2 - b_3^2)}{2\,a^2\,b^2} \,H_3 +
\frac{(a_2^2 - a_1^2)\,b_2\,b_3}{4\,a^2\,b^2} \,H_{Z_1}
- \frac{(2 a_3^2 - a_p^2)b_2\,b_3}{4\,a^2\,b^2} \,H_{Z_2}
+ \frac{(a_3^2 + a_1^2 -2 a_2^2)\,b_2\,b_3}{a^2\,b^2}\, H_m
\end{equation*}

\begin{equation*}
G_{45}=\frac{b_2\,b_3}{4\,b^2} \,\left[\,\frac{b_2^2 - 2 b_3^2}{b^2}\,(G_{Z_2} - G_3) + \frac{b_2^2}{b^2}\,(G_{Z_1} - G_3) - 4 \frac{(2\,b_2^2 - b_3^2)}{b^2} \,G_m \,\right]
\end{equation*}

\vspace{3mm}
\noindent {\bf Majorana L-handed, Dirac and Majorana R-handed:}

\begin{eqnarray}
N_{55}&=&F_{55}\,u_{11}^2 + 2\,q_{33}\,u_{11}\,u_{21} + G_{55}\,u_{21}^2  \\ \nonumber\\
N_{56}&=&F_{55}\,u_{11}\,u_{12} + q_{33}\,(u_{11}\,u_{22}+u_{12}\,u_{21}) + G_{55}\,u_{21}\,u_{22}  \\ \nonumber\\
N_{57}&=&F_{55}\,u_{11}\,u_{13} + q_{33}\,(u_{11}\,u_{23}+u_{13}\,u_{21}) + G_{55}\,u_{21}\,u_{23}  \\ \nonumber\\
N_{58}&=&F_{55}\,u_{11}\,u_{14} + q_{33}\,(u_{11}\,u_{24}+u_{14}\,u_{21}) + G_{55}\,u_{21}\,u_{24}  \\ \nonumber\\
N_{66}&=&F_{55}\,u_{12}^2 + 2\,q_{33}\,u_{12}\,u_{22} + G_{55}\,u_{22}^2  \\ \nonumber\\
N_{67}&=&F_{55}\,u_{12}\,u_{13} + q_{33}\,(u_{13}\,u_{22}+u_{12}\,u_{23}) + G_{55}\,u_{22}\,u_{23}  \\ \nonumber\\
N_{68}&=&F_{55}\,u_{12}\,u_{14} + q_{33}\,(u_{14}\,u_{22}+u_{12}\,u_{24}) + G_{55}\,u_{22}\,u_{24}  \\ \nonumber\\
N_{77}&=&F_{55}\,u_{13}^2 + 2\,q_{33}\,u_{13}\,u_{23} + G_{55}\,u_{23}^2  \\ \nonumber\\
N_{78}&=&F_{55}\,u_{13}\,u_{14} + q_{33}\,(u_{14}\,u_{23}+u_{13}\,u_{24}) + G_{55}\,u_{23}\,u_{24}  \\ \nonumber\\
N_{88}&=&F_{55}\,u_{14}^2 + 2\,q_{33}\,u_{14}\,u_{24} + G_{55}\,u_{24}^2                                                \end{eqnarray}

\begin{equation*}
F_{55}= \frac{a_1^2\,a_2^2}{a^4} \,F_1 + \frac{a_1^2\,a_3^2}{a^4} \,F_2 + \frac{a_2^2\,a_3^2}{a^4} \,F_3 +
\frac{(a_2^2 - a_1^2)^2}{4\,a^4} \,F_{Z_1}
+ \frac{(2 a_3^2 - a_p^2)^2}{12\,a^4} \,F_{Z_2}
+ \frac{(a_2^2 - a_1^2)\,(2\,a_3^2 - a_p^2)}{a^4}\, F_m
\end{equation*}

\begin{equation*}
q_{33}= \frac{a_2\,a_3\,b_2\,b_3}{a^2\,b^2} \,H_3 +
\frac{(a_2^2 - a_1^2)\,b_2^2}{4\,a^2\,b^2} \,H_{Z_1}
- \frac{(2 a_3^2 - a_p^2)\,(b_2^2 - 2 b_3^2)}{12\,a^2\,b^2} \,H_{Z_2}
+ \frac{a_3^2\,b_2^2 - a_p^2\,b_3^2  - a_2^2\,(b_2^2 - 2\,b_3^2)}{a^2\,b^2}\, H_m
\end{equation*}

\begin{equation*}
G_{55}= \frac{b_2^2\,b_3^2}{b^4} \,G_3 + \frac{b_2^4}{4\,b^4} \,G_{Z_1}
+ \frac{(b_2^2 - 2 b_3^2)^2}{12\,b^4} \,G_{Z_2} - \frac{b_2^2\,(b_2^2 - 2 b_3^2)}{b^4}\, G_m
\end{equation*}

\subsection{$( V_{CKM} )_{4\times 4}$ and $( V_{PMNS} )_{4\times 8}$  mixing matrices }

Within this $SU(3)$ family symmetry model, the transformation from
massless to physical mass fermion eigenfields for quarks and charged leptons is

\begin{equation*} \psi_L^o = V_L^{o} \:V^{(1)}_L \:\Psi_L \qquad \mbox{and}
\qquad \psi_R^o = V_R^{o} \:V^{(1)}_R \:\Psi_R \,,\end{equation*}

\noindent and for neutrinos $\Psi_\nu^o =  U_\nu^o \, U_\nu \,\Psi_\nu$. Recall now
that vector like quarks, Eq.(\ref{vectorquarks}), are $SU(2)_L$
weak singlets, and hence, they do not couple to $W$ boson in the
interaction basis. In this way, the interaction of  L-handed up and down quarks; ${f_{uL}^o}^T=(u^o,c^o,t^o)_L$ and
${f_{dL}^o}^T=(d^o,s^o,b^o)_L$, to the $W$ charged gauge boson is

\begin{equation} \frac{g}{\sqrt{2}} \,\bar{f^o}_{u L} \gamma_\mu f_{d L}^o
{W^+}^\mu = \frac{g}{\sqrt{2}} \,\bar{\Psi}_{u L}\;
[(V_{u L}^o\,V_{u L}^{(1)})_{3\times 4}]^T \;(V_{d L}^o\,V_{d L}^{(1)})_{3\times 4}\;
\gamma_\mu \Psi_{d L} \;{W^+}^\mu \:,\end{equation}

\noindent $g$ is the $SU(2)_L$ gauge coupling. Hence, the non-unitary $V_{CKM}$ of dimension $4\times
4$ is identified as

\begin{equation} (V_{CKM})_{4\times 4} = [(V_{u L}^o\,V_{u L}^{(1)})_{3\times 4}]^T \;(V_{d L}^o\,V_{d L}^{(1)})_{3\times 4}
\end{equation}

\noindent Similar analysis of the couplings of active L-handed neutrinos and L-handed charged leptons to $W$ boson,
leads to the lepton mixing matrix

\begin{equation} ( U_{PMNS} )_{4\times 8}   = [(V_{e L}^o\,V_{e L}^{(1)})_{3\times 4}]^T \;
(U_\nu^o\,U_\nu)_{3\times 8}
\end{equation}

\section{Numerical results} \label{numerical}

\emph{To illustrate the spectrum of masses and mixing, let us consider the following
fit of space parameters at the $M_Z$ scale \cite{xingzhang}}

\vspace{3mm}
Using the strong hierarchy for quarks and charged leptons masses\cite{prd2007}, here
we report the fermion masses and mixing, coming out from a global fit of the parameter space.

\vspace{3mm}
In the approach $\alpha_2 \approx \alpha_3 = \alpha_H$, we take the input values

\begin{equation*} M_1 = 10\,\text{TeV} \quad , \quad M_2 = 1\,\text{TeV} \quad , \quad \frac{\alpha_H}{\pi}=0.05 \end{equation*}

\noindent for the $M_1$, $M_2$ horizontal boson masses, Eq.(\ref{M1M2}), and the $SU(3)$  coupling
constant, respectively, and the ratio of electroweak VEV's: $V_i$ from $\Phi^d$, and  $v_i$ from $\Phi^u$

\begin{equation*} \frac{V_1}{V_2} = 0.09981 \quad , \quad \frac{\sqrt{V_1^2+V_2^2}}{V_3} = 0.54326
\end{equation*}

\begin{equation*} \frac{v_1}{v_2} = 0.1 \quad , \quad \frac{\sqrt{v_1^2+v_2^2}}{v_3} = 0.5
\end{equation*}

\subsection{Quark masses and mixing}

\noindent {\bf u-quarks:}

\vspace{3mm}
\noindent Tree level see-saw mass matrix:

\begin{equation} {\cal M}_u^o=
\begin{pmatrix}
 0 & 0 & 0 & 7933.76 \\
 0 & 0 & 0 & 79337.6 \\
 0 & 0 & 0 & 159467. \\
 0 & 1.18613\times 10^6 & -841128. & 374542.
\end{pmatrix}\,\text{MeV} \,,\end{equation}

\noindent the mass matrix up to one loop corrections:

\begin{equation} {\cal M}_u=
\begin{pmatrix}
 -1.40509 & 187.442 & -66.8139 & -255.74 \\
 -0.125675 & -609.844 & 408.793 & 1564.71 \\
 -0.062809 & -1197.67 & -172100. & 1825.79 \\
 -0.001885 & -35.9461 & 14.3165 & 1.502\times 10^6
 \end{pmatrix}\,\text{MeV} \end{equation}

\noindent and the u-quark masses

\begin{equation}
(\,m_u \;,\; m_c \;,\; m_t \;,\; M_U\,)=
(\,1.3802 \;,\; 640.801 \;,\;172105\;,\;1.502\times 10^6\,)\,\text{MeV}
\end{equation}

\vspace{5mm}
\noindent {\bf d-quarks:}

\vspace{3mm}
\begin{equation} {\cal M}_d^o=
\begin{pmatrix}
 0 & 0 & 0 & 1740.94 \\
 0 & 0 & 0 & 17442.3 \\
 0 & 0 & 0 & 32265.8 \\
 0 & 70019.9 & -41383.4 & 910004
\end{pmatrix}\;\text{MeV} \end{equation}

\vspace{3mm}
\begin{equation} {\cal M}_d=
\begin{pmatrix}
 3.09609 & 28.1593 & -47.4565 & -4.23475 \\
 0.271539 & -40.5966 & 215.617 & 19.2404 \\
 0.147401 & -176.235 & -2846.26 & 37.484 \\
 0.005900 & -7.05504 & 16.8159 & 914365.
\end{pmatrix}\;\text{MeV} \end{equation}

\vspace{3mm}
\begin{equation}
(\,m_d \;,\; m_s \;,\; m_b \;,\; M_D\,)=
(\, 2.82  \;,\; 61.9998 \;,\; 2860   \;,\; 914365 \,)\;\text{MeV}
\end{equation}

\noindent and the quark mixing

\begin{equation} V_{CKM}=
\begin{pmatrix}
 0.974352 & 0.225001 & 0.003647 & 0.000410 \\
 -0.224958 & 0.973502 & 0.041031 & -0.001417 \\
 -0.005632 & 0.040662 & -0.997868 & -0.039994 \\
 0.000576 & -0.002325 & 0.031130 & 0.001251
\end{pmatrix}  \label{vckm} \end{equation}

\vspace{5mm}
\subsection{Charged leptons:}

\vspace{3mm}
\begin{equation} {\cal M}_e^o=
\begin{pmatrix}
 0 & 0 & 0 & 28340.3 \\
 0 & 0 & 0 & 283940. \\
 0 & 0 & 0 & 525249. \\
 0 & 17105.4 & -11570.9 & 5.94752\times 10^6
\end{pmatrix}\;\text{MeV} \end{equation}

\vspace{3mm}
\begin{equation} {\cal M}_e=
\begin{pmatrix}
 -0.499137 & 29.7086 & -43.9181 & -0.15097 \\\\
 -0.043776 & -72.8148 & 238.953 & 0.821414 \\\\
 -0.023663 & -183.913 & -1720.65 & 1.18425 \\\\
 -0.002378 & -18.4839 & 34.6241 & 5.977\times 10^6
\end{pmatrix}\;\text{MeV} \end{equation}

\noindent fit the charged lepton masses:

\begin{equation*}
( m_e \,,\, m_\mu \,,\, m_\tau \,,\, M_E ) = ( 0.486 \,,\,102.7\,,\,1746.17\,,\, 5.977\times 10^6\, )\,\text{MeV}
\end{equation*}

\noindent and the mixing

\begin{equation} V_{e \,L}^o\, V_{e \,L}^{(1)}=
\begin{pmatrix}
 0.968866 & 0.24054 & -0.0584594 & 0.00474112 \\
 0.205175 & -0.912554 & -0.350561 & 0.0475013 \\
 -0.138557 & 0.330471 & -0.929446 & 0.0878703 \\
 -0.00217545 & 0.0132348 & 0.0990967 & 0.994987
\end{pmatrix} \end{equation}

\vspace{3mm}
\subsection{Neutrino masses and mixing:}

\begin{equation} \mathcal{M}_\nu^o=\, \text{eV}
\begin{pmatrix}
 0 & 0 & 0 & 0 & 0 & 0 & 53594.6 & 44137.2 \\
 0 & 0 & 0 & 0 & 0 & 0 & 535946. & 441372. \\
 0 & 0 & 0 & 0 & 0 & 0 & 1.07\times 10^6 & 887147. \\
 0 & 0 & 0 & 0 & 0 & 0 & 0 & 0 \\
 0 & 0 & 0 & 0 & 0 & 0 & 1.80\times 10^6 & 1.49\times 10^6 \\
 0 & 0 & 0 & 0 & 0 & 0 & -886604. & -730152. \\
 53594.6 & 535946. & 1.07\times 10^6 & 0 & 1.8097\times 10^6 & -886604. & 1.97\times 10^8 & 4.88\times 10^8 \\
 44137.2 & 441372. & 887147. & 0 & 1.49\times 10^6 & -730152. & 4.88\times 10^8 & 7.02\times 10^8
\end{pmatrix} \end{equation}

\vspace{3mm}
\begin{equation} \mathcal{M}_\nu=\, \text{eV}
\begin{pmatrix}
 -0.0119 & 0.0527 & 0.0227 & -0.0878 & -0.0693 & 0.1674 & -0.0016 & 0.0004 \\
 0.0527 & -0.036 & 0.002 & 0.068 & 0.043 & -0.748 & 0.007 & -0.002 \\
 0.0227 & 0.002 & 0. & 0. & 0.0008 & 0.0005 & -5.2\times 10^{-6} & 1.5\times 10^{-6}\\
 -0.0878 & 0.068 & 0. & -0.125 & -0.1218 & 1.282 & -0.012 & 0.003 \\
 -0.0693 & 0.043 & 0.0008 & -0.121 & 3.206 & -0.7430 & 0.0074 & -0.0021 \\
 0.1674 & -0.748 & 0.0005 & 1.282 & -0.7430 & 1749.96 & 0.0003 & -0.0001 \\
 -0.0016 & 0.007 & -5.2\times 10^{-6} & -0.012 & 0.0074 & 0.0003 & -1.\times 10^8 & 1.1\times 10^{-6} \\
 0.0004 & -0.002 & 1.5\times 10^{-6} & 0.003 & -0.0021 & -0.0001 & 1.1\times 10^{-6} & 1.\times 10^9
\end{pmatrix}
\end{equation}

\noindent generates the neutrino mass eigenvalues

\begin{equation}
(m_1, m_2, m_3, m_4, m_5, m_6, m_7, m_8)=  \text{eV}\,
(\,0,\, -0.0085\,,\, 0.049\,,\, -0.22\,,\, 3.21\,,\, 1749.96\,,\, -1\times 10^8\,,\,
 1\times 10^9\,) \,,\label{neutrinomasses}
\end{equation}

\noindent the squared mass differences

\begin{equation}
m_2^2-m_1^2 \approx 0.0000723 \, \text{eV}^2\quad , \quad m_3^2-m_1^2 \approx 0.0024 \, \text{eV}^2
\end{equation}

\begin{equation}
m_4^2-m_1^2 \approx 0.0492 \, \text{eV}^2 \quad , \quad m_5^2-m_1^2 \approx 10.3182\, \text{eV}^2
\end{equation}

\noindent and the lepton mixing matrix

\begin{equation} U_{PMNS}=
\begin{pmatrix}
 0.2104 & 0.3520 & 0.8658 & -0.2861 & 0.0060 & 0.0053 & 0.00005 & 0.00001 \\
 -0.8282 & 0.0030 & 0.0186 & -0.5478 & 0.1038 & -0.0507 & -0.0005 & -0.0001 \\
 0.0807 & 0.0041 & 0.0052 & 0.0881 & 0.8475 & -0.5074 & -0.0050 & -0.0014 \\
 0.0034 & 0.0003 & 0.0011 & -0.0021 & -0.0857 & 0.0512 & 0.0005 & 0.0001
\end{pmatrix} \end{equation}

\section{Conclusions}

We have reported a low energy parameter space, within a local $SU(3)$
Family symmetry model, which combines tree level "Dirac
See-saw" mechanisms and radiative corrections to implement a
successful hierarchical spectrum, for charged fermion masses and quark mixing.
In section \ref{numerical} we illustrated the predicted values for quark and charged
lepton masses at the the $M_Z$ scale\cite{xingzhang},
and a non-unitary quark mixing matrix $(V_{CKM})_{4\times 4}$ within allowed
values reported in PDG 2012 \cite{PDG2012}, coming from a
parameter space with the horizontal gauge boson masses within (1-10) TeV, the
$SU(2)_L$ weak singlet vector-like fermion masses
$M_D \approx 914.365 \,\text{GeV}$, $M_U \approx 1.5 \,\text{TeV}$,
$M_E \approx 5.97 \,\text{TeV}$, the neutrino masses in  Eq.(\ref{neutrinomasses}), including
two light sterile neutrinos, and the squared neutrino mass differences:
$m_2^2-m_1^2 \approx 7.23 \times 10^{-5}\, \text{eV}^2$, $m_3^2-m_1^2 \approx 2.4 \times 10^{-3}\, \text{eV}^2$,
$m_4^2-m_1^2 \approx 0.049\, \text{eV}^2$, $m_5^2-m_1^2 \approx 10.3\, \text{eV}^2$. Therefore, the new
particles introduced in this model are within reach at the current LHC and neutrino oscillation experiments.

\vspace{3mm}
\emph{It is worth to comment here that the symmetries and the transformation of the
fermion and scalar fields, all together, forbid tree level Yukawa couplings between
ordinary standard model fermions. Consequently, the flavon scalar fields introduced to break
the symmetries: $\Phi^u$, $\Phi^d$, $\eta_2$ and $\eta_3$, couple only
ordinary fermions to their corresponding vector like fermion at tree level. Thus, FCNC
scalar couplings to ordinary fermions are suppressed by light-heavy mixing angles,
which as is shown in $(V_{CKM})_{4 \times 4}$, Eq.(\ref{vckm}), may be
small enough to suppress properly the FCNC mediated by the scalar fields within this scenario.}

\vspace{5mm}
\section*{Acknowledgements}

It is my pleasure to thank the organizers N.S. Mankoc-Borstnik, H.B. Nielsen, M. Y. Khlopov,
and participants for the stimulating Workshop at Bled, Slovenia. This work was
partially supported by the "Instituto Polit\'ecnico Nacional",
(Grants from EDI and COFAA) and "Sistema Nacional de
Investigadores" (SNI) in Mexico.



\begin{thebibliography}{99}

\bibitem{Altarelli.Smirnov} G. Altarelli, \emph{An Overview of Neutrino Mixing},
arXiv: 1210.3467 [hep-ph]; A. Yu Smirnov, \emph{Neutrino 2012: Outlook - Theory},
arXiv: 1210.4061 [hep-ph];

\bibitem{DayaBay} DAYA-BAY Collaboration, F. P. An \emph{et. al., Observation of
Electron-Antineutrino  Disappearance at the Daya Bay}, arXiv: 1203.1669.

\bibitem{T2K} T2K Collaboration, K. Abe \emph{et. al., Indication of Electron Neutrino
Appearance from an Accelerator-Produced Off-Axis Muon Neutrino Beam},
Phys. Rev. Lett. {\bf 107}, 041801(2011).

\bibitem{MINOS} MINOS Collaboration, P. Adamson \emph{et. al., Improved search for muon-neutrino
to electron-neutrino oscillations in MINOS},
Phys. Rev. Lett. {\bf 107}, 181802(2011).

\bibitem{DOUBLE} DOUBLE-CHOOZ Collaboration, Y. Abe \emph{et. al., Indication for the Disappearance
of Reactor Electron Antineutrinos in the Double Chooz Experiment}, arXiv: 1207.6632.

\bibitem{RENO} RENO Collaboration, J. K. Ahn \emph{et. al., Observation of Reactor
Electron Antineutrino Disappearance in the Reno Experiment},arXiv: 1204.0626.

\bibitem{TBM} F. Harrison, D. H. Perkins and W. G. Scott, {\cal Phys. lett. B}{\bf 530}, 167 (2002);
See also: R. Gaitan, A. Hernandez-Galeana, J.M.
Rivera-Rebolledo and P. Fernandez de Cordoba,
"Neutrino mixing and masses in a left-right model with mirror fermions", European Physical Journal C {\bf 72} , (2012) 1859-1866 ; arXiv:1201.3155[hep-ph]

\bibitem{MiniBooNE} {\bf MiniBooNE} Collaboration, A. A. Aguilar-Arevalo \emph{et. al.,
 A Combined $\nu_\mu \rightarrow \nu_e$ and $\bar{\nu}_\mu \rightarrow \bar{\nu}_e$
 Oscillation Analysis of the  MiniBooNE Excesses.}, arXiv: 1207.4809.

\bibitem{LSND-MiniBooNe} J.M. Conrad, W.C. Louis, and  M.H. Shaevitz, arXiv:1306.6494;
I. Girardi, A. Meroni, and S.T. Petcov, arXiv:1308.5802; M. Laveder and C. Giunti, arXiv:1310.7478;
A. Palazzo, arXiv:1302.1102; O. Yasuda, arXiv:1211.7175;
J. Kopp, M. Maltoni and T. Schwetz, Phys. Rev. Lett.
{\bf 107}, (2011)  091801; C. Giunti, arXiv:1111.1069; 1107.1452; F. Halzen,
arXiv:1111.0918; Wei-Shu Hou and Fei-Fan Lee, arXiv:1004.2359; O. Yasuda, arXiv:1110.2579;
Y.F. Li and Si-shuo Liu, arXiv:1110.5795; B. Bhattacharya, A. M. Thalapillil, and C. E. M. Wagner,
arXiv:1111.4225; J. Barry, W. Rodejohann and He Zhang, arXiv:1110.6382[hep-ph]; JHEP {\bf 1107},(2011)091; F.R. Klinkhamer, arXiv:1111.4931[hep-ph].

\bibitem{earlyradm} S. Weinberg, Phys. Rev. Lett.{\bf
29}, 388(1972); H. Georgi and S.L. Glashow, Phys. Rev. {\bf D 7},
2457(1973); R.N. Mohapatra, Phys. Rev. {\bf D 9}, 3461(1974); S.M.
Barr and A. Zee, Phys. Rev. {\bf D 15}, 2652(1977); H. Georgi,
"Fermion Masses in Unified models", in {\it First Workshop in
Grand Unification}, ed. P.H. Frampton, S.L. Glashow, and A.
Yildiz(1980, Math Sci Press, Brookline, MA); S.M. Barr ,Phys. Rev.
{\bf D 21}, 1424(1980); R. Barbieri and D.V. Nanopoulos, Phys.
Lett. {\bf B 95}, 43(1980); S.M. Barr ,Phys. Rev. {\bf D 24},
1895(1981); L.E. Ibanez, Phys. Lett. {\bf B 177}, 403(1982); B.S.
Balakrishna, A.L. Kagan and R.N. Mohapatra, Phys. Lett. {\bf B
205}, 345(1988); B.S. Balakrishna, Phys. Rev. Lett. {\bf 60},
1602(1988); K.S. Babu and E. Ma, Mod. Phys. Lett. {\bf A 4 },
1975(1989); H.P. Nilles, M. Olechowski and S. Pokorski, Phys.
Lett. {\bf B 248}, 378(1990); R. Rattazzi, Z. Phys.{\bf C 52},
575(1991); K.S. Babu and R.N. Mohapatra, Phys. Rev. Lett. {\bf
64}, 2747(1990); X.G. He, R.R. Volkas, and D.D. Wu, Phys. Rev.
{\bf D 41}, 1630(1990); Ernest Ma, Phys. Rev. Lett. 64 (1990)
2866. B.A. Dobrescu and P.J. Fox, JHEP {\bf 0808}, 100(2008); S.M.
Barr, Phys. Rev. {\bf D 76}, 105024(2007); S.M. Barr and A. Khan,
Phys. Rev. {\bf D 79}, 115005(2009);

\bibitem{modeldiscrete}
Sandip Pakvasa and Hirotaka Sugawara, Phys. Lett. {\bf B 73}
61(1978); Y. Yamanaka, H. Sugawara, and S. Pakvasa, Phys. Rev.
{\bf D 25} 1895(1982); K. S. Babu and X.G. He, ibid. 36
3484(1987); Ernest Ma, Phys. Rev. Lett. {\bf B 62} 61(1989).

\bibitem{modelcontinuous}
A. Davidson, M. Koca, and K. C. Wali, Phys. Rev. Lett. {\bf B 43}
92(1979), Phys. Rev. {\bf D 20} 1195(1979); C. D. Froggatt and H.
B. Nielsen, Nucl. Phys. {\bf B 147} 277(1979); A. Sirlin, Phys.
Rev. {\bf D 22} 971(1980); A. Davidson and K. C. Wali, ibid. 21
787(1980).

\bibitem{modelrad}
X.G. He, R. R: Volkas, and D. D. Wu, Phys. Rev. {\bf D 41}
1630(1990); Ernest Ma, Phys. Rev. Lett. {\bf 64} 2866(1990).

\bibitem{medscalars}
E. Garcia, A. Hernandez-Galeana, D. Jaramillo, W. A. Ponce and A.
Zepeda, Revista Mexicana de Fisica Vol. {\bf 48(1)}, 32(2002); E.
Garcia, A. Hernandez-Galeana, A. Vargas and A. Zepeda,
hep-ph/0203249.

\bibitem{prd2007} A.Hernandez-Galeana, Phys.Rev. {\bf D
76}, 093006(2007).

\bibitem{DSB} N. Chen, T. A. Ryttov, and R. Shrock, arXiv:1010.3736 [hep-ph];
C. T. Hill and E. H. Simmons,
Phys. Rept. {\bf 381}, 235 (2003); {\it Workshop on Dynamical Electroweak
Symmetry Breaking}, Southern Denmark Univ. 2008 (http://hep.sdu.dk/dewsb);
R.S. Chivukula, M. Narain, and J. Womersley, in Particle Data Group, J.
Phys. G {\bf 37} 1340, (2010); F. Sannino, Acta Phys. Polon. B {\bf 40},
3533 (2009)(arXiv:0911.0931).

\bibitem{albinosu32004} A. Hernandez-Galeana, Rev. Mex. Fis. {\bf Vol. 50(5)},
(2004) 522. hep-ph/0406315.

\bibitem{su3models} For some references on $SU(3)$ family symmetry see: \\
J.L. Chkareuli, C.D. Froggatt, and H.B. Nielsen, Nucl. Phys. B {\bf 626},
(2002) 307;T. Appelquist, Y. Bai, and M. Piai, Phys. Lett. B {\bf 637}, 245
(2006); T. Appelquist, Y. Bai, and M. Piai, Phys. Rev. D {\bf 74},
076001 (2006), and references therein.

\bibitem{albinosu3bled} A. Hernandez-Galeana, Bled Workshops in Physics, (ISSN:1580-4992),{\bf Vol. 13, No. 2}, (2012) Pag. 28; arXiv:1212.4571[hep-ph]; {\bf Vol. 12, No. 2}, (2011) Pag. 41; arXiv:1111.7286[hep-ph]; {\bf Vol. 11, No. 2}, (2010) Pag. 60; arXiv:1012.0224[hep-ph];  Bled Workshops in Physics,{\bf Vol. 10, No. 2}, (2009) Pag. 67; arXiv:0912.4532[hep-ph];

\bibitem{SU3MKhlopov}
Z.G.Berezhiani and M.Yu.Khlopov, {\it Sov.J.Nucl.Phys.} 51 (1990) 739; 935; {\it Sov.J.Nucl.Phys.} 52 (1990) 60; {\it Z.Phys.C- Particles and Fields} 49 (1991) 73;
Z.G.Berezhiani, M.Yu.Khlopov and R.R.Khomeriki, {\it Sov.J.Nucl.Phys.} 52 (1990) 344;
A.S.Sakharov and M.Yu.Khlopov {\it Phys.Atom.Nucl.} 57 (1994) 651;
M.Yu. Khlopov:
\emph{Cosmoparticle physics}, World Scientific, New York
-London-Hong Kong - Singapore, 1999; M.Yu. Khlopov:
\emph{Fundamentals of Cosmoparticle physics}, CISP-Springer, Cambridge, 2011;
Z.G. Berezhiani, J.K. Chkareuli, {\em JETP Lett.} {\bf 35} (612) 1982; {\em JETP Lett.} {\bf 37} (338) 1983; Z.G. Berezhiani, {\em Phys. Lett. B} {\bf 129} (99) 1983.

\bibitem{normaapproach} N.S. Mankoc-Borstnik, arXiv: 1011.5765; Bled Workshops in Physics,{\bf Vol. 12, No. 2}, (2011) Pag. 112; A. Hernandez-Galeana and N.S. Mankoc-Borstnik,Bled Workshops in Physics,{\bf Vol. 12, No. 2}, (2011) Pag. 55: arXiv:1112.4368[hep-ph]; {\bf Vol. 11, No. 2}, (2010) Pag. 89 , Pag. 105; G. Bregar, R.F. Lang and N.S. Mankoc-Borstnik, Pag.161; M. Y. Khlopov and  N.S. Mankoc-Borstnik, Pag.177; arXiv:1012.0224[hep-ph]; N.S. Mankoc-Borstnik, Bled Workshops in Physics,
{\bf Vol. 10, No. 2}, (2009) Pag. 119; G. Bregar and N.S. Mankoc-Borstnik, Pag. 149; arXiv:0912.4532[hep-ph]

\bibitem{khlopov} M. Yu Khlopov, A. G. Mayorov, and E. Yu Soldatov, Bled Workshops in Physics,  {\bf Vol. 12, No. 2}, (2011) Pag. 94; arXiv:1111.3577[hep-ph];
{\bf Vol. 11, No. 2}, (2010) Pag. 73; Pag. 185; arXiv:1012.0224[hep-ph]; {\bf Vol. 10, No. 2}, (2009) Pag. 79; M.Y. Khlopov, Pag. 155; arXiv:0912.4532[hep-ph];

\bibitem{vectorlikepapers} J.A. Aguilar-Saavedra, R. Benbrik, S. Heinemeyer, and
M. P\'erez-Victoria, arXiv:1306.0572; J.A. Aguilar-Saavedra, arXiv:1306.4432; Jonathan M. Arnold,
Bartosz Fornal and Michael Trott,  JHEP 1008:059, 2010, arXiv:1005.2185 and references
therein.

\bibitem{T.Yanagida1979} T. Yanagida, Phys. Rev. D {\bf 20}, 2986
(1979).

\bibitem{ATLAS} G. Aad \emph{et. al.}, ATLAS Collaboration, Phys. Lett. {\bf B 716}, 1(2012),
arXiv: 1207.7214.

\bibitem{CMS} S. Chatrchyan \emph{et. al.}, CMS Collaboration, Phys. Lett. {\bf B 716}, 30(2012),
arXiv: 1207.7235.

\bibitem{xingzhang} Zhi-zhong Xing, He Zhang and Shun Zhou, Phys. Rev. D {\bf 86}, 013013 (2012).

\bibitem{PDG2012} J. Beringer {\it et al.}, ( Particle Data Group ),
Phys. Rev. D {\bf 86}, 010001 (2012).

\end{thebibliography}
\end{document}